\documentclass[twocolumn]{raa}
\usepackage{caption,subcaption}
\usepackage{graphicx,times}          
\usepackage{natbib}
\usepackage{overpic}
\usepackage{ulem} 
\usepackage{amssymb,amsmath}
\usepackage{float}
\usepackage{multirow}

\usepackage[a4paper=true,pagebackref=true]{hyperref}
\hypersetup{pdftitle = The title of my PDF, pdfauthor = My name, pdfsubject= The subject, pdfkeywords = keyword1 keyword2 keyword3} 
\hypersetup{colorlinks = true, linkcolor = green, anchorcolor = red, citecolor = blue, filecolor = red, pagecolor = red, urlcolor = red}

\def\ls{\lower 2pt \hbox{$\;\scriptscriptstyle \buildrel<\over\sim\;$}}

\begin{document}

    \title{Hydrodynamical Simulations of the Triggering of Nuclear Activities by Minor Mergers of Galaxies}
    
    \volnopage{ {\bf 20XX} Vol.\ {\bf X} No. {\bf XX}, 000--000}
    \setcounter{page}{1}

    \author{Chao Yang\inst{1,2,3}, Junqiang Ge\inst{1,3}, Youjun Lu\inst{1,2,3}}

    \institute{National Astronomical Observatories, Chinese Academy of Sciences, 20A Datun Road, Beijing 100101, China;{\it chyang@nao.cas.cn, luyj@nao.cas.cn}\\
    \and
    School of Astronomy and Space Sciences, University of Chinese Academy of Sciences, 19A Yuquan Road, Beijing 100049, China\\
    \and
    CAS Key laboratory for computational Astrophysics, National Astronomical Observatories, Chinese Academy of Sciences, Beijing, 100012, China\\
    \vs \no
    {\small Received 20XX Month Day; accepted 20XX Month Day}
    }

\abstract{
Major mergers of galaxies are considered to be an efficient way to trigger Active Galactic Nuclei and are thought to be responsible for the phenomenon of quasars. This has however recently been challenged by observations of a large number of low luminosity Active Galactic Nuclei at low redshift ($z\lesssim1$) without obvious major merger signatures. Minor mergers are frequently proposed to explain the existence of these Active Galactic Nuclei. In this paper, we perform nine high resolution hydrodynamical simulations of minor galaxy mergers and investigate whether nuclear activities can be efficiently triggered by minor mergers, by setting various properties for the progenitor galaxies of those mergers. We find that minor galaxy mergers can activate the massive black hole in the primary galaxy with an Eddington ratio of $f_{\rm Edd}>0.01$ and $>0.05$ (or a bolometric luminosity $>10^{43}$ and $>10^{44}\mathrm{erg\, s^{-1}}$) with a duration of $2.71$ and $0.49$\,Gyr (or $2.69$ and $0.19$\,Gyr), respectively. The nuclear activity of primary galaxy strongly depends on the nucleus separation, the nucleus is more active as the two nuclei approach to each other. Dual Active Galactic Nuclei systems can still possibly form by minor mergers of galaxies, the time period for dual Active Galactic Nuclei is only $\sim 0.011$ Gyr and $\sim 0.017$ Gyr with Eddington ratio of $f_{\rm Edd}>0.05$ and bolometric luminosity $>10^{44}\mathrm{erg\, s^{-1}}$. This time period is typically shorter than that of dual Active Galactic Nuclei induced by galaxy major mergers.
}

   \authorrunning{C. Yang et al. }            
   \titlerunning{Nuclear activity triggered by galaxy minor merger}  
   \maketitle

\section{Introduction}

Galaxy merger is a key process for the formation and evolution of galaxies in the $\Lambda$CDM hierarchical structure formation paradigm. It is widely accepted that the existence of massive black holes (MBH) in the centers of nearby galaxies is ubiquitous \citep{2013ARA&A..51..511K, 1995ARA&A..33..581K}. During the merging of two galaxies, an massive binary black hole (BBH) system may form naturally since the two central MBHs approach each other due to the dynamical friction and viscous drag \citep{1980Natur.287..307B, 2002MNRAS.331..935Y}. In the meantime, the angular momentum of gas may be also transferred out due to tidal interactions during the merging process, which leads to the gas sinking into the vicinity of one MBH or both MBHs to trigger nuclear activity or activities \citep[e.g.,][]{ 1989Natur.340..687H}. If only one of the MBH is activated, then the system may appear as an offset Active Galactic Nucleus (offset AGN , or oAGN), and if both of the two MBHs are activated, the system may appear as a dual AGN (dAGN) \citep[e.g.,][]{2012ApJ...748L...7V, 2012ApJ...753...42C, 2014ApJ...789..112C, 2015ApJ...806..219C, 2016ApJ...829...37B, 2016ApJ...830...50M, 2017ApJ...838..129B, 2017MNRAS.469.4437C, 2017ApJ...847...41C, 2018MNRAS.478.3056B}.

Galaxy mergers may provide an efficient way to transfer the gas angular momentum outward and lead to the sinking of gas into the vicinity of MBHs and thus trigger nuclear activities. However, the connection between galaxy merger and AGN triggering is still observationally inconclusive. Some observations clearly show the disky structure of the low redshift low luminosity AGN host galaxies, which lead to the proposal that galaxy minor merger or secular processes triggers AGN activity \citep{2011ApJ...726...57C, 2012ApJ...744..148K, 2017MNRAS.470..755H, 2017MNRAS.465.2895L, 2019MNRAS.483.2441V}. While some other observations find that AGN host galaxies are highly perturbed in morphology, which lead to a claim that galaxy major merger dominates the triggering of AGNs \citep{2011MNRAS.418.2043E, 2012ApJ...758L..39T, 2014A&A...569A..37M, 2014MNRAS.441.1297S, 2015ApJ...804...34H, 2018ApJ...853...63D, 2018PASJ...70S..37G}. These two lines of observational results apparently contradict with each other and hinder the understanding of the triggering of nuclear activities.

Both dAGNs and oAGNs can be used as tracers of galaxy mergers. According to hydrodynamical simulations if both progenitor galaxies are gas rich and comparable in mass (mass ratio $> 1/3$, i.e., a major merger), both MBHs may be activated with relatively large Eddington ratios and high bolometric luminosities, and they emerge as a dAGN system with the two nuclei separated on a scale of $\sim 1-10$\,kpc \citep[e.g.,][]{2012ApJ...748L...7V, 2013MNRAS.429.2594B, 2015MNRAS.447.2123C, 2016MNRAS.458.1013S, 2017MNRAS.469.4437C}. 

Observations do find such dAGN systems by using different techniques \citep[e.g.,][]{2003ApJ...582L..15K, 2004ApJ...604L..33Z, 2009ApJ...705L..76W, 2009ApJ...705L..20X, 2009ApJ...698..956C, 2010ApJ...708..427L, 2010ApJ...715L..30L, 2011ApJ...737L..19C, 2011ApJ...733..103F, 2011ApJ...740L..44F, 2011ApJ...735L..42K, 2012MNRAS.425.1185F, 2012ApJ...745...67F, 2012ApJS..201...31G, 2012ApJ...746L..22K, 2013MNRAS.429.2594B, 2015ApJ...813..103M, 2016MNRAS.457.3878Z, 2018ApJ...867...66C, 2019MNRAS.482.1889W}.
 
The conditions for the formation of oAGNs may be different from dAGNs as the activation of only one MBH is required. Both major and minor mergers may be responsible for the formation of dAGN, but it is still not clear which one dominates the contribution to oAGNs \citep[e.g.][]{2015ApJ...806..219C, 2018ApJ...869..154B}.\footnote{We also note here that in the final stage of a galaxy merger, the merged MBH may gain a recoiling speed up to several thousand $\mathrm{km\ s^{-1}}$ due to the asymmetric gravitational wave emission \citep{2007PhRvL..98w1102C}. The recoiled BH can carry almost everything bounded to it (within $\sim 10^{5}$ gravitational influence radius), as a consequence, the broad line region (BLR) will run away bounded to the central MBH and the narrow line region (NLR) will be left behind \citep{2004ApJ...606L..17M, 2011MNRAS.412.2154B, 2016ApJ...829...37B, 2018MNRAS.475.5179S}. This may also contribute to the census of oAGNs significantly.}  

In this paper, we use high resolution hydrodynamical simulations to study whether significant nuclear activities can be triggered by minor galaxy mergers and investigate whether dAGNs and oAGNs can emerge from the merging processes of galaxy minor mergers. The paper is organized as follow. In section~\ref{sec:method}, we briefly introduce the numerical simulations we performed and their initial setups. Then we summarize the main results in Section~\ref{sec:result}. Finally, conclusions and discussions are given in Section~\ref{sec:conclusion}.

\section{Numerical simulations}
\label{sec:method}

\subsection{Initial setup}

We use the smoothed particle hydrodynamics (SPH) code GADGET-2 \citep{2005MNRAS.364.1105S} for simulation, in which we take into account those physical processes including the star formation, supernova feedback, BH accretion, and AGN feedback.

To simulate the star formation and supernova feedback processes, the gas density ($\rho_{\rm gas}$) is divided into a hot component $\rho_{\mathrm{h}}$ and a cold component $\rho_{\mathrm{c}}$ based on the hybrid model proposed in \cite{2003MNRAS.339..289S}, from which we have $\rho_{\rm gas} = \rho_{\mathrm{h}} + \rho_{\mathrm{c}}$. The star formation rate at a characteristic timescale $t_{*}$ is defined as
\begin{equation}
\frac{\mathrm{d}\rho_{*}}{\mathrm{d}t} = (1 - \beta)\frac{\rho_{\mathrm{c}}}{t_{*}}
\end{equation}
where $\beta$ denote the mass fraction of the newly formed stars exploded as the supernovae instantly. A gas particle will spawn a star particle when the gas density exceeds a given threshold $\rho_{\mathrm{th}} = 0.35\ h^{2}\ \mathrm{cm}^{-3}$ to match the observational law \citep{2003MNRAS.339..289S, 2004MNRAS.348..435N, 2014ApJ...780..145T}.

With the defined gas density, the MBH accretion rate is calculated by adopting the Bondi-Hoyle-Lyttleton parametrization \citep{1939PCPS...35..405H, 1944MNRAS.104..273B, 1952MNRAS.112..195B} formalism:
\begin{equation}
\dot{M} = \frac{4 \pi \alpha G^{2} M_{\mathrm{BH}}^{2} \rho_{\rm gas}}{(c_{\mathrm{s}}^{2} + v^{2})^{3/2}}
\end{equation}
where $\alpha$ is a dimensionless parameter and is set as $\alpha = 8$ for $z = 3$ cases, the same as that adopted in the literature, e.g., \citet{2009MNRAS.398...53B},\citet{ 2009ApJ...690..802J}, \citet{2019SCPMA}, and $\alpha = 100$ for $z = 1$ cases to consider the mass resolution of $z=1$ cases is $10$ times lower than that of the $z=3$ cases \citep[for discussions on the settings of $\alpha$, see][]{2005MNRAS.361..776S, 2014MNRAS.442.1992H, 2015MNRAS.454.1038R, 2017MNRAS.467.3475N}\footnote{We choose a higher booster factor $\alpha = 100$ to calculate the Bondi accretion rate for those three $z=1$ cases. However, we find that the accretion rate is still relatively low compared to $z = 3$ cases. An adaptive mesh refinement (AMR) method or a $10$ times higher mass resolution may be required to better understand the gas feeding to the very central region of the galaxy. We defer this to a future study.}, $c_{\mathrm{s}}$ corresponds to the sound speed of the gas, $v$ is the velocity of the BH relative to the gas. These settings on $\alpha$ ensure a reasonable BH accretion rate. Here we limit $\dot{M}$ to be not larger than the Eddington accretion rate $\dot{M}_{\mathrm{Edd}}$, in order to avoid super-Eddington accretion.

As to the AGN feedback, the energy injected into the surrounding gas is a fraction ($\epsilon_{\mathrm{f}}$) of the AGN bolometric luminosity 
\begin{equation}
\dot{E}_{\mathrm{feed}} = \epsilon_{\mathrm{f}} L_{\rm bol} = \epsilon_{\mathrm{f}} \epsilon_{\mathrm{r}} \dot{M} c^{2}
\end{equation}
where $c$ is the light speed in the vacuum, $\epsilon_{\mathrm{r}}$ is the mass to energy converting efficiency. Here we take the typical values $\epsilon_{\mathrm{r}} = 0.1$ and $\epsilon_{\mathrm{f}} = 0.05$ to study the AGN feedback \citep{2005Natur.433..604D}, which can regulate the evolution of the established galaxy following the observed $M_{\mathrm{BH}}-\sigma$ relation \citep[e.g.,][]{1998AJ....115.2285M, 2002ApJ...574..740T, 2013ARA&A..51..511K}. In the simulation, the BH accretion process and AGN feedback are numerically implemented by following the procedure described in \cite{2005MNRAS.361..776S}.

\subsection{Construction of galaxy merger systems}

To study whether and how the central MBH can be triggered by minor mergers, we design 9 sets of these systems with different mass ratios and galaxy types as listed in Table \ref{tab:ini_setup}. We denote those simulations according to their parameter settings on $(\textsf{q}, \textsf{f}, \textsf{z})$ in the following way (see Table \ref{tab:ini_setup}). Here \textsf{q} represents the mass ratio of two progenitor galaxies, \textsf{q5} and \textsf{q10} represent $1:5$ and $1:10$ minor mergers, respectively; \textsf{f} marks the gas fraction of each galaxy in the unit of $0.1$, the two consequent numbers after \textsf{f} represent the gas fraction of the primary galaxy and secondary galaxy, respectively, e.g., \textsf{f13} denotes the gas fractions of the primary galaxy and secondary galaxy are $0.1$ and $0.3$, respectively; \textsf{ss} and \textsf{es} represent the type of the primary and secondary galaxies, and \textsf{s} and \textsf{e} represent spiral galaxy and elliptical galaxy, respectively; \textsf{z1} and \textsf{z3} represent the initial redshift of the simulation, i.e. $z = 1$ and $z = 3$, respectively. One of these nine cases has an extra symbol \textsf{p10}, which means that we specify a pericenter $r_{\rm p} = 10$\,kpc, for other cases we adopt the pericenter as the $20$\% of the virial radius of the primary galaxy (9.3 kpc for $z = 3$ and 35.7 kpc for $z = 1$). 

At redshift $z=3$, galaxies tend to be gas rich and possibly have varied gas fractions. We hence start the simulation from \textsf{q5f13ssz3}, \textsf{q5f31ssz3}, and \textsf{q5f33ssz3}, which are 1:5 mergers started from $z=3$ but with different gas fractions included in the primary and secondary galaxies, with which we can quantify the  effect of gas content on the AGN triggering.  These three mergers are set to be co-planar (with the inclination angle $i=0^{\circ}$) and prograde. Considering that different inclination angles, which indicate different angular momentum of these merging systems and can affect the morphology of the merged galaxy \citep[e.g.,][]{2005ApJ...622L...9S, 2017MNRAS.470.3946S, 2019SCPMA}, we then set the \textsf{q5f33ssz3i} system with inclination angle $i=45^{\circ}$ but keep other parameters the same as \textsf{q5f33ssz3} to investigate the inclination angle effect. We also set the galaxy mergers with mass ratio 1:10 with $i=0^{\circ}$ (\textsf{q10f33ssz3}) and $i=45^{\circ}$ (\textsf{q10f33ssz3i}) to analyze how the triggering of AGN activity is affected by different mass ratios. The above 6 galaxy mergers are put into a parabolic Keplerian orbit (eccentricity $e = 1$), with the initial separation set as the sum of the virial radii of the primary and secondary galaxies. And the pericenter is set to 20\% of the virial radius of the primary galaxy.

At redshift $z=1$, the fraction of elliptical galaxies increases compared to that at $z=3$ \citep[e.g.,][]{2008ApJS..175..356H, 2010ApJ...709..644I, 2014ARA&A..52..291C}. We then simulate one spiral-spiral (\textsf{q5f33ssz1}) and one elliptical-spiral (\textsf{q5f03esz1}) merging systems started at $z=1$ to specify their difference with that happened in $z=3$. These two systems have the same orbital setup as that started at $z=3$. In addition, we set a elliptical-spiral minor merger (\textsf{q5f03esz1p10}) with lower pericenter $r_{\mathrm{p}} = 10\  \mathrm{kpc}$ to make a closer encounter at the first pericentric passage to identify whether the gas transfer can be significantly changed in the merging process. 

In our simulation, a spiral galaxy consists of a dark matter halo, a stellar bulge, a disk component with both stars and gas included, and a central MBH. An elliptical galaxy includes a dark matter halo, a stellar bulge, and a central MBH. 

For the spiral galaxy, we use a Hernquist profile \citep{1990ApJ...356..359H} to describe its dark matter halo with the virial mass $M_{\mathrm{vir}}$ given in Table \ref{tab:ini_galaxy}. The disk mass is set as 0.04 of the virial mass, i.e. $m_{\mathrm{d}} = 0.04 M_{\mathrm{vir}}$. Inside the disk, the gas fraction $f_{\mathrm{gas}}$ varies from 0.1 to 0.3 for spiral galaxies. The galaxy bulge, which is also assumed to distribute as a Hernquist profile, is set as 0.008 of the virial mass, which indicates an initial bulge-to-total ratio B/T=0.2. According to the $M_{\mathrm{BH}}$-$M_{\mathrm{Bulge}}$ relation \citep[e.g.,][]{2003ApJ...589L..21M}, we set the central MBH has a mass fraction $m_{\mathrm{BH}} = 1.0875\times 10^{-5}$ of the virial mass, to guarantee the establishment of a typical and reasonable spiral galaxy. The MBH particle is settled down in the galactic center as a sink particle, which accretes the surrounding gas particles including both the mass and momentum. 

For the elliptical galaxy, the disk and gas components are excluded. Both the bulge and dark matter halo are described by the Hernquist profile. 
The bulge mass fraction is $m_{\mathrm{b}} = 0.05$, and the BH mass fraction is $m_{\mathrm{BH}} = 8.0 \times 10^{-5}$, which is also set based on the $M_{\mathrm{BH}}$-$M_{\mathrm{Bulge}}$ relation.

All the other parameter setup of the elliptical and spiral galaxies are listed in Table \ref{tab:ini_galaxy}, and the corresponding mass resolution and softening lengths of the four particles: dark matter, bulge, disk, and gas are listed in Table \ref{tab:ini_resolution}.

\begin{table*}
\centering
\caption{Physical parameters for constructed galaxy mergers}
\begin{tabular}{ccccccccc}
\hline
Simulation & Galaxy Type &$M_{\mathrm{vir}1}(M_{\odot})$ & $M_{\mathrm{vir}2}(M_{\odot})$ & q & $f_{\mathrm{gas}1}$ & $f_{\mathrm{gas}2}$ & $z$ & Notes\\
\hline
\textsf{q5f13ssz3}  & spiral + spiral & $2.27\times 10^{11}$ & $4.54\times 10^{10}$ & 1:5 & 0.1 & 0.3 & 3 & ... \\
\textsf{q5f31ssz3}  & spiral + spiral & $2.27\times 10^{11}$ & $4.54\times 10^{10}$ & 1:5 & 0.3 & 0.1 & 3 & ... \\
\textsf{q5f33ssz3}  & spiral + spiral & $2.27\times 10^{11}$ & $4.54\times 10^{10}$ & 1:5 & 0.3 & 0.3 & 3 & ... \\
\textsf{q5f33ssz3i} & spiral + spiral & $2.27\times 10^{11}$ & $4.54\times 10^{10}$ & 1:5 & 0.3 & 0.3 & 3 & Inclined by $45^{\circ}$ \\
\hline
\textsf{q10f33ssz3}  & spiral + spiral & $2.27\times 10^{11}$ & $2.27\times 10^{10}$ & 1:10 & 0.3 & 0.3 & 3 & ... \\
\textsf{q10f33ssz3i} & spiral + spiral & $2.27\times 10^{11}$ & $2.27\times 10^{10}$ & 1:10 & 0.3 & 0.3 & 3 & Inclined by $45^{\circ}$ \\
\hline
\textsf{q5f33ssz1}    & spiral + spiral & $2.0\times 10^{12}$ & $4.0\times 10^{11}$ & 1:5 & 0.3 & 0.3 & 1 & ...\\
\textsf{q5f03esz1}    & elliptical + spiral & $2.0\times 10^{12}$ & $4.0\times 10^{11}$ & 1:5 & 0 & 0.3 & 1 & ... \\
\textsf{q5f03esz1p10} & elliptical + spiral & $2.0\times 10^{12}$ & $4.0\times 10^{11}$ & 1:5 & 0 & 0.3 & 1 & $r_{p} = 10$\, kpc \\
\hline
\end{tabular}
\label{tab:ini_setup}
\end{table*}

\begin{table*}
\caption{Physical parameters of individual galaxies in our simulation.}
\begin{center}
\begin{tabular}{c|ccc|ccc}
\hline
\multicolumn{1}{c}{} & \multicolumn{3}{|c|}{$z = 1$} & \multicolumn{3}{c}{$z = 3$}\\
\cline{2-7}
Symbol & Primary(E) & Primary(S) & Secondary(S) & Primary & Secondary(1:5) & Secondary(1:10)\\
(1) & (2) & (3) & (4) & (5) & (6) & (7)\\
\hline
$M_{\mathrm{vir}}\ (M_{\odot})$ & $2.0 \times 10^{12}$ & $2.0 \times 10^{12}$ & $4.0 \times 10^{11}$ & $ 2.3 \times 10^{11}$ & $4.5 \times 10^{10}$ & $2.3 \times 10^{10}$\\
$m_{\mathrm{d}}\ (M_{\mathrm{vir}})$ & 0 & 0.04 & 0.04 & 0.04 & 0.04 & 0.04\\
$m_{\mathrm{b}}\ (M_{\mathrm{vir}})$ & 0.05 & 0.008 & 0.008 & 0.008 & 0.008 & 0.008\\
$M_{\mathrm{BH}}\ (M_{\odot})$ & $1.6 \times 10^{8}$ & $2.2 \times 10^{7}$ & $4.4 \times 10^{6}$ & $3.0 \times 10^{6}$ & $6.0 \times 10^{5}$ & $3.0 \times 10^{5}$\\

$j_{\mathrm{d}}\ (j_{\mathrm{halo}})$ & 0 & 0.04 & 0.04 & 0.04 & 0.04 & 0.04\\
$R_{200}\ (\mathrm{kpc})$ & 178.50 & 178.50 & 104.42 & 46.53 & 27.08 & 21.59\\
$R_{\mathrm{H}}\ (\mathrm{kpc})$ & 67.12 & 67.12 & 39.26 & 8.66 & 5.04 & 4.02\\
$R_{\mathrm{S}}\ (\mathrm{kpc})$ & 59.50 & 59.50 & 34.80 & 5.17 & 3.01 & 2.40\\
$H\ (\mathrm{kpc})$ & 0 & 5.73 & 3.35 & 0.94 & 0.55 & 0.44\\
$Z_{0}\ (\mathrm{kpc})$ & 0 & 0.57 & 0.34 & 0.10 & 0.05 & 0.04\\
$A\ (\mathrm{kpc})$ & 2.78 & 1.15 & 0.67 & 0.19 & 0.11 & 0.09\\
\hline
\end{tabular}
\end{center}
\label{tab:ini_galaxy}
Note: The left column shows symbols used for describing a galaxy, which from top to bottom represent the virial mass, disk mass fraction, bulge mass fraction, BH mass, disk spin, halo virial radius, scale radius of Hernquist profile, halo scale radius, disk scale length, disk thickness, and bulge scale radius, respectively. Column (2)-(7) list the correspond values, (E) stands for the elliptical galaxy, (S) means the spiral one, (1:5) and (1:10) represent the secondary spiral galaxies in 1:5 and 1:10 minor mergers, respectively. 
\end{table*}

\begin{table*}
\centering
\caption{Mass and spatial resolutions for different particles at $z = 1$ and $z = 3$ } 
\begin{tabular}{c|cc|cc}
\hline
 & \multicolumn{2}{|c|}{Mass Resolution ($M_{\odot}$)} & \multicolumn{2}{c}{Softening Length (pc)}\\
\cline{2-5}
Particle Type & $z = 1$ & $z = 3$ & $z = 1$ & $z = 3$\\
\hline
Dark Matter & $1.1 \times 10^6$  & $1.1 \times 10^5$ & 30 & 30 \\
Bulge & $3.7 \times 10^{4}$  & $3.7 \times 10^{3}$ & 10 & 10\\
Disk & $3.7 \times 10^{4}$  & $3.7 \times 10^{3}$ & 10 & 10\\
Gas & $4.6 \times 10^{4}$  & $4.6 \times 10^{3}$ & 20 & 20\\
\hline
\end{tabular}
\label{tab:ini_resolution}
\end{table*}

\subsection{Identification of AGN activities}

In the galaxy merging process, the gas can be concentrated to the galaxy center and trigger the nuclear activity, but the corresponding detection depends on the detectability of current telescopes. Therefore, in this paper, we set two thresholds for the bolometric luminosity ($L_{\rm bol}=10^{43} {\rm erg~s^{-1}}$, $L_{\rm bol}=10^{44} {\rm erg~s^{-1}}$), and two in Eddington ratio ($f_{\rm Edd}=0.01$, $f_{\rm Edd}=0.05$) to match with varied detection capabilities of different telescopes.

\section{Results}
\label{sec:result}

\begin{figure*}
    \centering
    \includegraphics[scale=0.43]{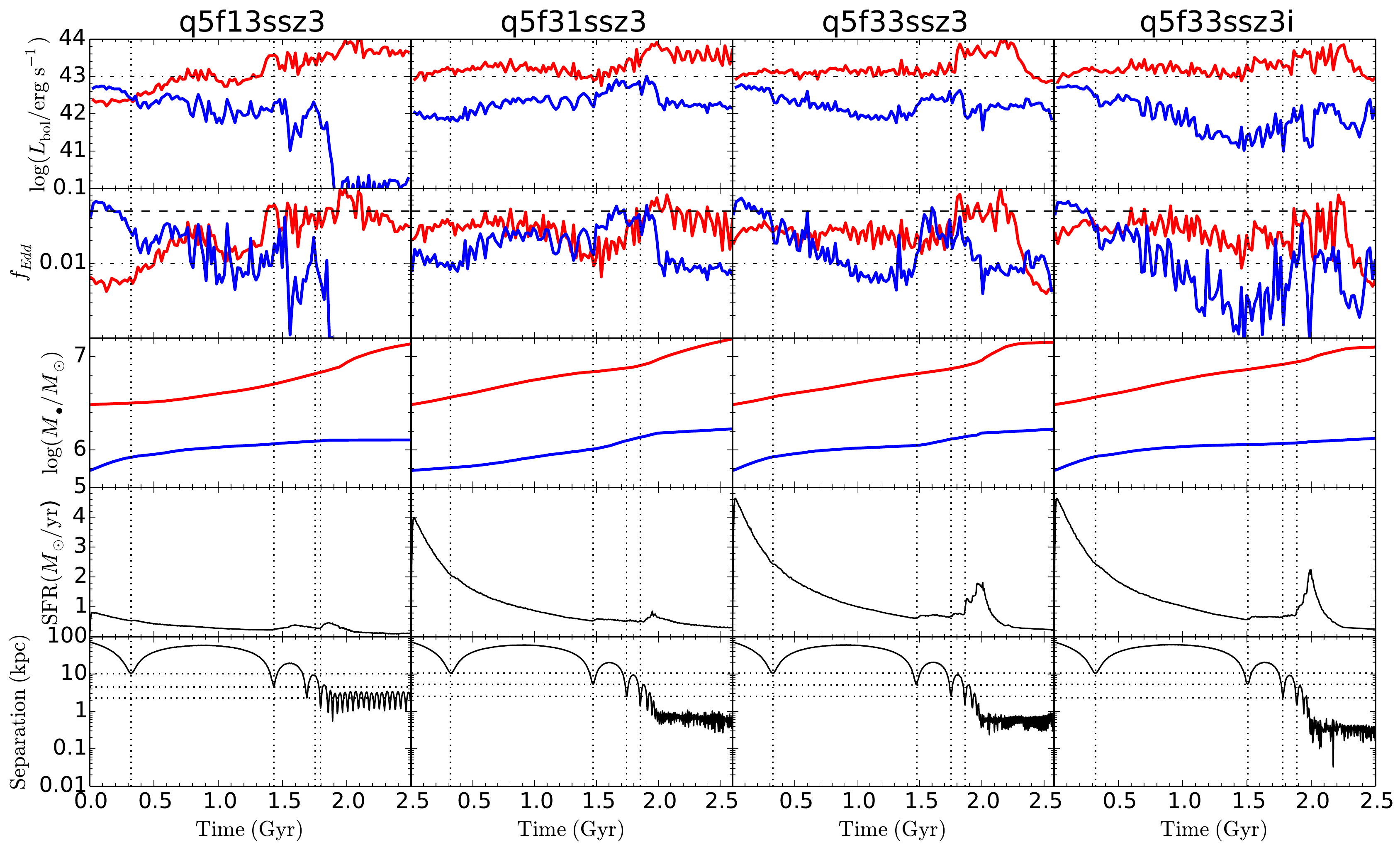}
    \caption{Evolution of the nuclear bolometric luminosities (first row), Eddington ratios (second row), MBH masses (third row), total star formation rate (fourth row), and the separation of the primary and secondary MBHs (fifth row). Columns from left to right show the results obtained from the four simulations starting from $z = 3$ with mass ratio of $1:5$. The red and blue solid lines in the first to third rows represent the corresponding evolution curves for the primary and secondary MBHs, respectively. In each column, the vertical dotted lines from left to right mark the cosmic time at the first, second, third, and fourth pericentric passages, respectively. The three horizontal dotted lines in the bottom row indicate separations of the two MBHs at the first to third pericentric passages, respectively.}
    \label{fig:bhacc_z3_5}
\end{figure*}

\begin{figure*}
    \centering
    \includegraphics[scale=0.43]{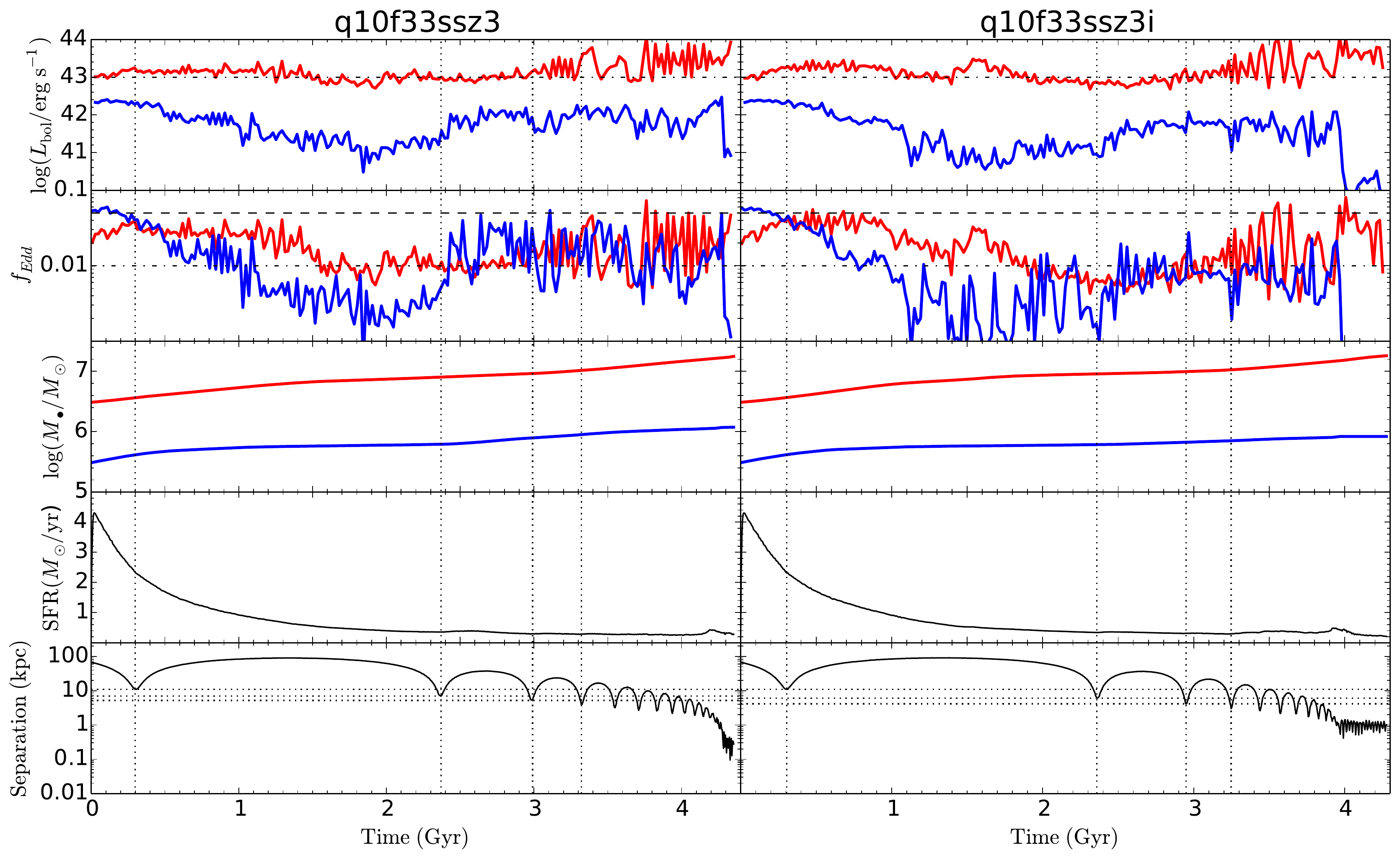}
    \caption{Parameter evolution of the minor mergers started at $z=3$ with mass ratio 1:10 (\textsf{q10f33ssz3} in left and \textsf{q10f33ssz3i} in right). Lines and colors are the same as shown in Figure~\ref{fig:bhacc_z3_5}.}
    \label{fig:bhacc_z3_10}
\end{figure*}

\begin{figure*}
    \centering
    \includegraphics[scale=0.43]{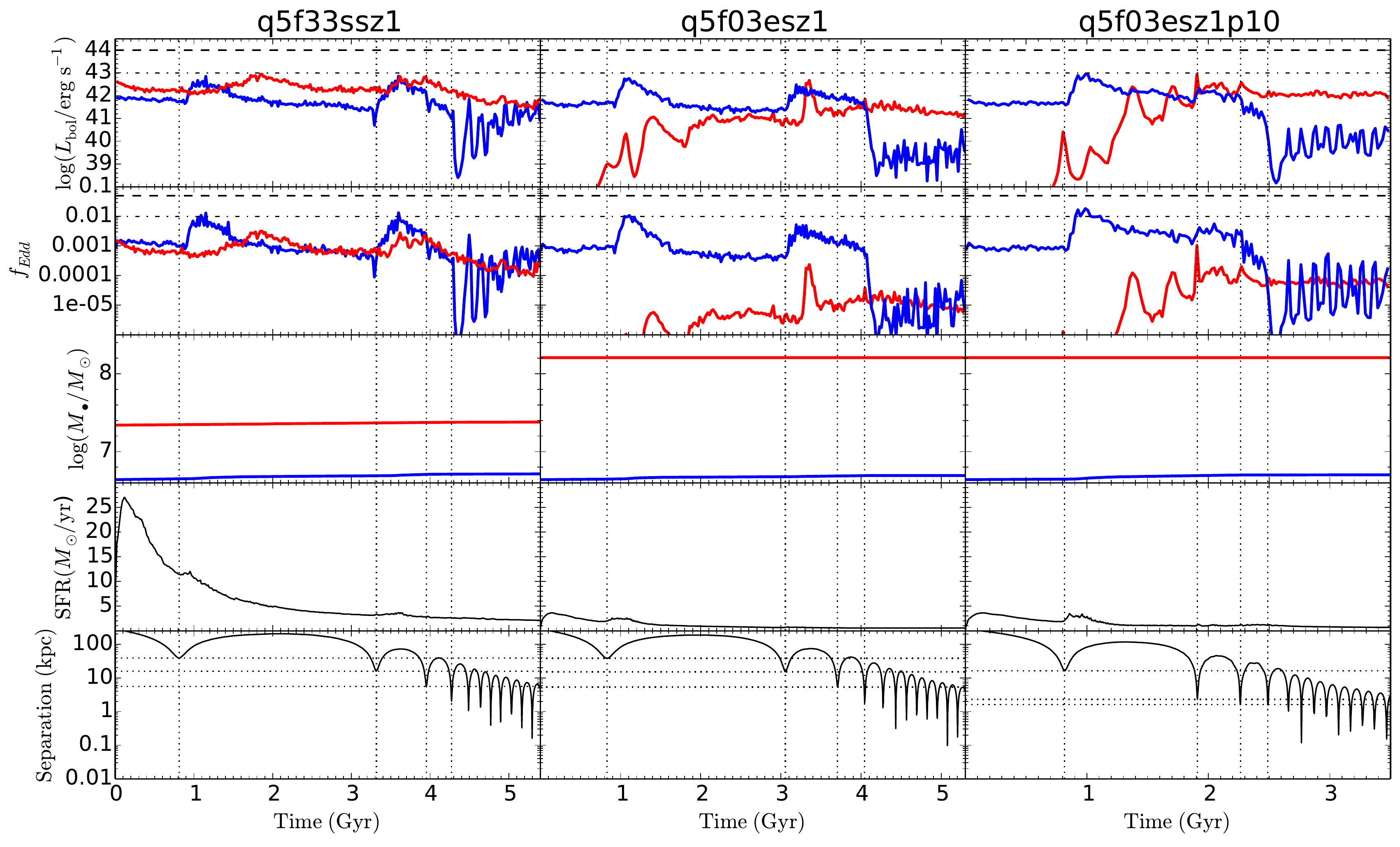}
\caption{Parameter evolution of the minor mergers started at $z=1$ with mass ratio 1:5. Columns from left to right show the evolution curves of \textsf{q5f33ssz1}, \textsf{q5f03esz1}, and \textsf{q5f03esz1p10} simulations, respectively. Lines and colors are the same as shown in Figure~\ref{fig:bhacc_z3_5}.}
    \label{fig:bhacc_z1}
\end{figure*}

Figures~\ref{fig:bhacc_z3_5}, \ref{fig:bhacc_z3_10}, and \ref{fig:bhacc_z1} show the evolution processes of the nuclear bolometric luminosity, Eddington ratio, BH mass, star formation rate (SFR), and BH separation of all the 9 minor mergers as listed in Table \ref{tab:ini_setup}. The nine sets of evolution present us how the primary and secondary galaxies play their roles in the minor merger.

Based on our constructed galaxy mergers, we can have a clear understanding on how the orbital decay rely on different initial conditions. The orbital decay of the first three simulations in Figures~\ref{fig:bhacc_z3_5} and the first two simulations in Figure \ref{fig:bhacc_z1} are quite similar before the first three pericentric passages, which indicates that the dynamical friction at the early stage of minor mergers is determined by the mass ratio, instead of gas content \citep{2002MNRAS.331..935Y}. Those minor mergers with an inclination angle (last column of Figure \ref{fig:bhacc_z3_5} and second column of Figure \ref{fig:bhacc_z3_10}) or smaller $r_{\rm P}$ (last two columns of Figure \ref{fig:bhacc_z1}) can accelerate the merging process . When the mass ratio decrease from $1:5$ (Figure \ref{fig:bhacc_z3_5}) to $1:10$ (Figure \ref{fig:bhacc_z3_10}), the merging time increases by a factor of $\sim 2$ which can be easily understood as the dynamical friction timescale is inversely proportional to the mass of the secondary galaxy.

At the beginning of the minor merger, the bolometric luminosities and Eddington ratios of the primary and secondary MBHs are determined by their initial gas fraction. For the \textsf{q5f13ssz3}, \textsf{q5f03esz1} and \textsf{q5f03esz1p10} simulations, the gas fraction of the primary galaxy is lower than the secondary one, which causes correspondingly lower $L_{\rm bol}$ and $f_{\rm Edd}$. For the other six cases, the primary MBHs still have lower $f_{\rm Edd}$ but their $L_{\rm bol}$ are higher than the secondary galaxy at most of the evolution time. 

Once the merging two galaxies go through the first pericentric passage, the primary galaxy begins to rob the gas from the secondary galaxy, and the bolometric luminosity of the primary MBH is systematically larger than the secondary galaxy for those gas-rich mergers. For the two elliptical-spiral galaxy minor mergers (right two columns of Figure~\ref{fig:bhacc_z1}), the evidence of the gas capture is more clear: both the $L_{\rm bol}$ and $f_{\rm Edd}$ of the primary MBH increase dramatically after the first pericenter, and their $f_{\rm Edd}$ are comparable with the secondary MBH after the fourth pericentric passage, which means their $L_{\rm bol}$ are $\sim 5$ times higher than the secondary MBH. This gas capture process can actually decrease the nuclear activity of the secondary MBH. On the other hand, the tidal torques can also enhance the gas concentration to the secondary MBH. In all the nine cases we can see that both the $L_{\rm bol}$ and $f_{\rm Edd}$ increase after the first to fourth pericentric passages as shown by the four vertical dotted lines in each panel. The gas capture and tidal torque finally produce the oscillated $L_{\rm bol}$ and $f_{\rm Edd}$ evolution curves.

Due to the similar evolution processes, the MBH masses of the two galaxies increase in similar trends for those minor mergers started at $z=3$ (Figures \ref{fig:bhacc_z3_5} and \ref{fig:bhacc_z3_10}). The primary MBH increase about $\sim 0.6$ dex, while the secondary MBH only have a maximum increase of $\sim 0.2$ dex. For those galaxy mergers started at $z=1$ (Figure~\ref{fig:bhacc_z1}), since their galaxy and MBH masses are 10 times larger than that at $z=3$, minor merger can not supply enough gas accretion and the MBH masses increase less than $0.1$ dex.

The amplitude of the SFR evolution for the 9 galaxy mergers are different, and determined by the total amount of gas included in the two galaxies. The total gas mass included in the \textsf{q5f13ssz3} merger (left column of Figure~\ref{fig:bhacc_z3_5}) is weaker than others because of its smallest amount of gas fraction included in the two galaxies. On the contrary, the \textsf{q5f33ssz1} (left column of Figure~\ref{fig:bhacc_z1}) contains the largest amount of gas, which then have the strongest SFR.

\begin{figure*}
    \centering
    \includegraphics[width=0.9\textwidth]{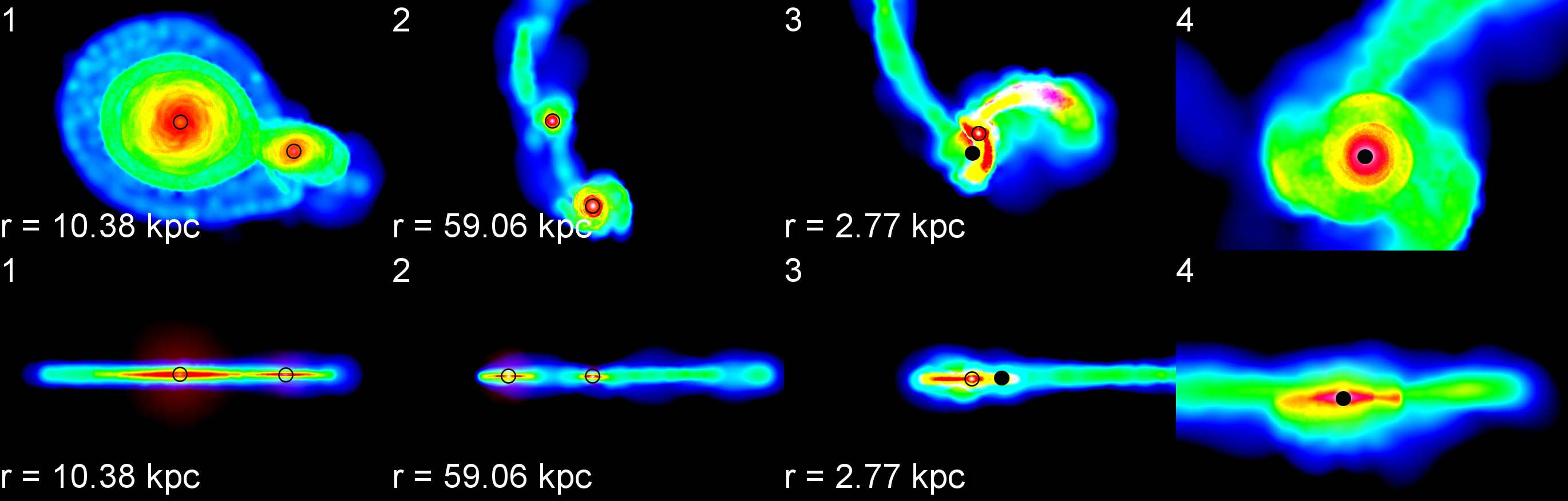}\\        
    \includegraphics[width=0.9\textwidth]{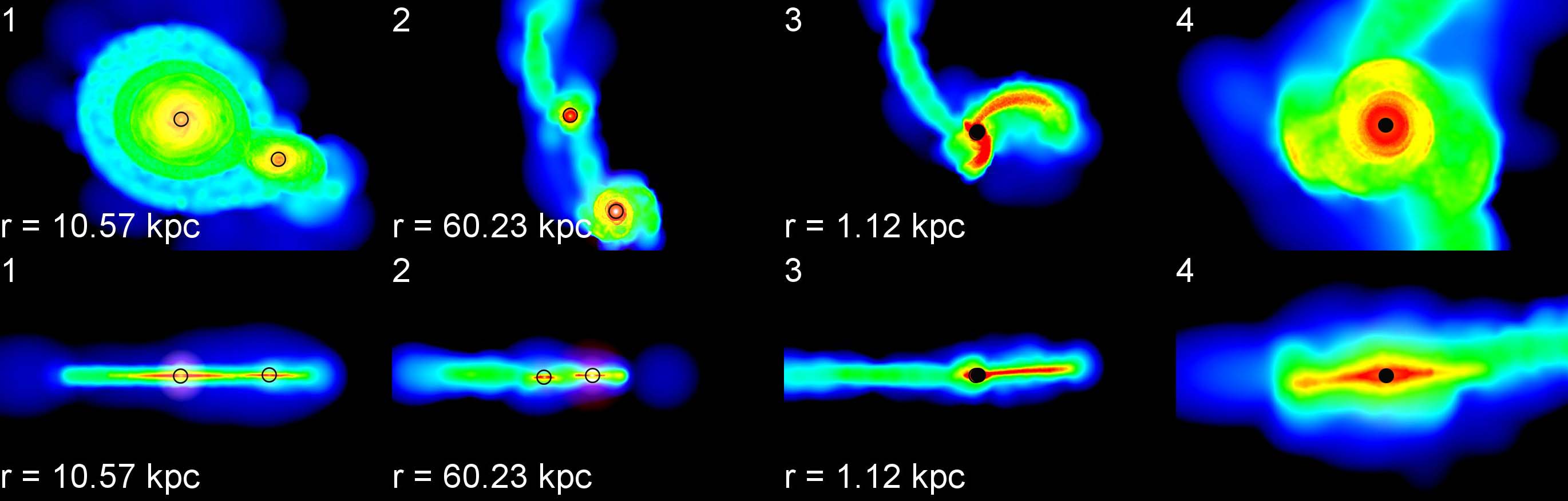}\\
    \includegraphics[width=0.9\textwidth]{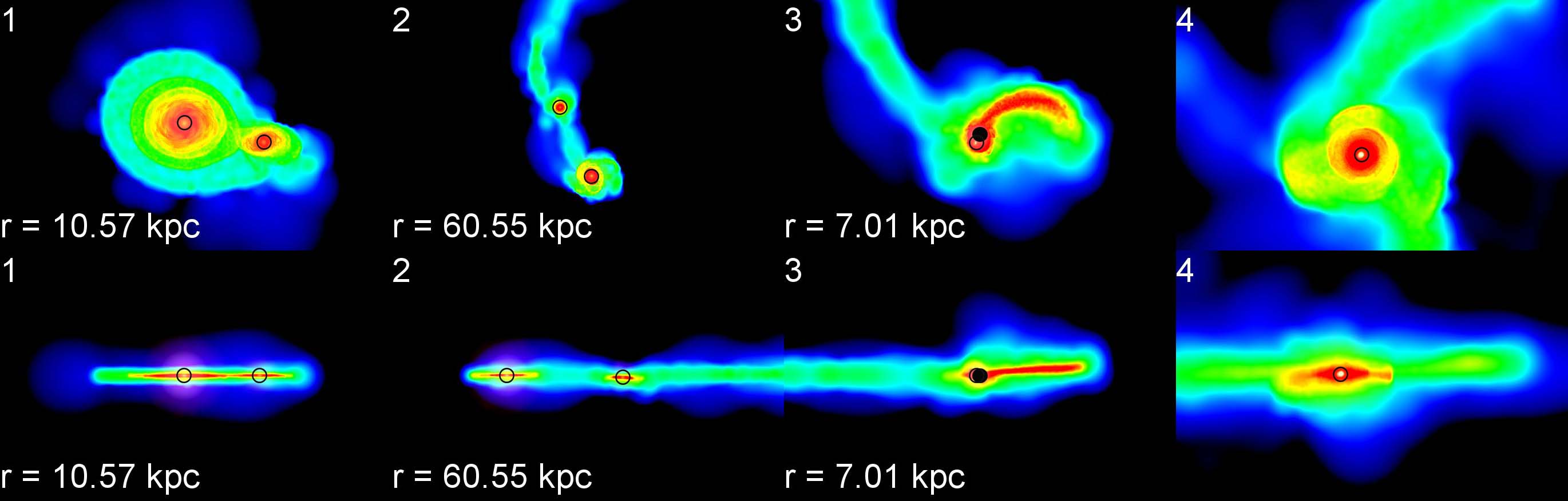}\\
    \includegraphics[width=0.9\textwidth]{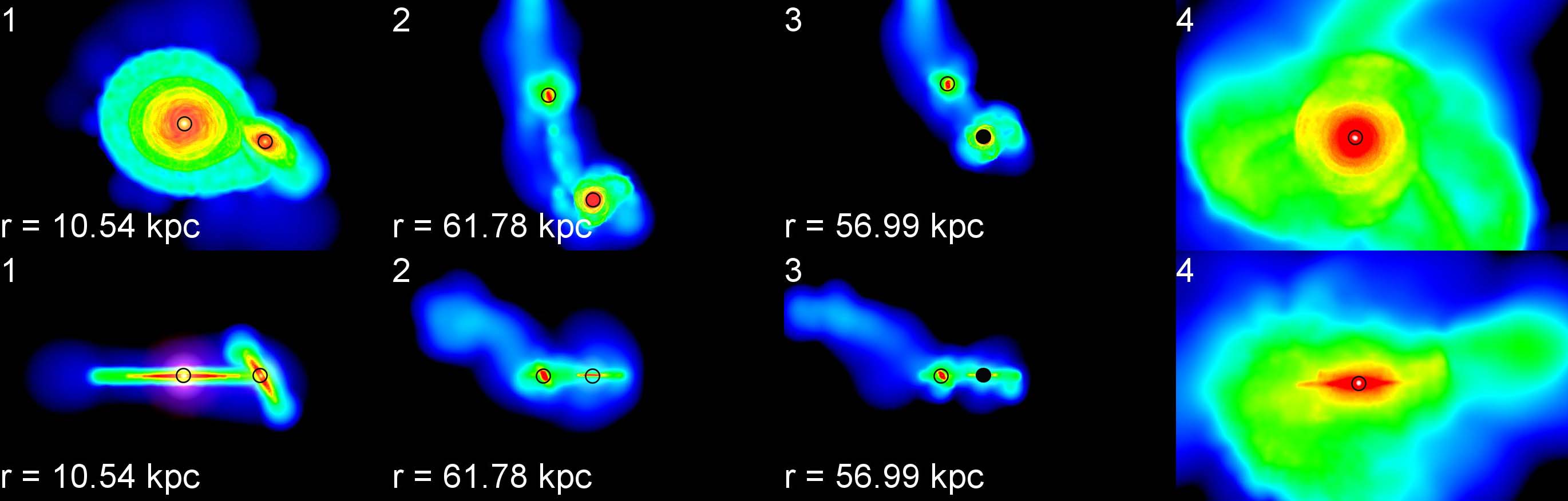}\\
    \caption{Snapshots of the four minor mergers started at $z = 3$ with mass ratio 1:5. Every two rows from top to bottom correspond to the simulations of \textsf{q5f13ssz3}, \textsf{q5f31ssz3}, \textsf{q5f33ssz3}, and \textsf{q5f33ssz3i}. In each two rows, the first row shows the four snapshots viewed in face-on (perpendicular to the galactic plane of the primary galaxy), and the second row shows that viewed in edge-on (parallel to the galactic plane of the primary galaxy). Numbers 1-4 at the top left of each panel represent the four snapshots during the merger: (1) the first pericentric passage, (2) the first apocentric passage after the first pericentric passage, (3) one of the two nuclei is active, and (4) the last output of the simulation. The separation between the two MBHs is given at the bottom left of each panel. The black circles in each panel show the position of the two MBHs, those MBHs with $L_{\rm bol}>10^{43}{\rm erg~ s^{-1}}$ are shown in filled black circle, while those MBHs with $L_{\rm bol}<10^{43}{\rm erg~ s^{-1}}$ are shown in open black circles. The radii of the circles are not scaled to the real size of MBHs.}
    \label{fig:galaxy1}
\end{figure*}

\begin{figure*}
    \centering
    \includegraphics[width=0.9\textwidth]{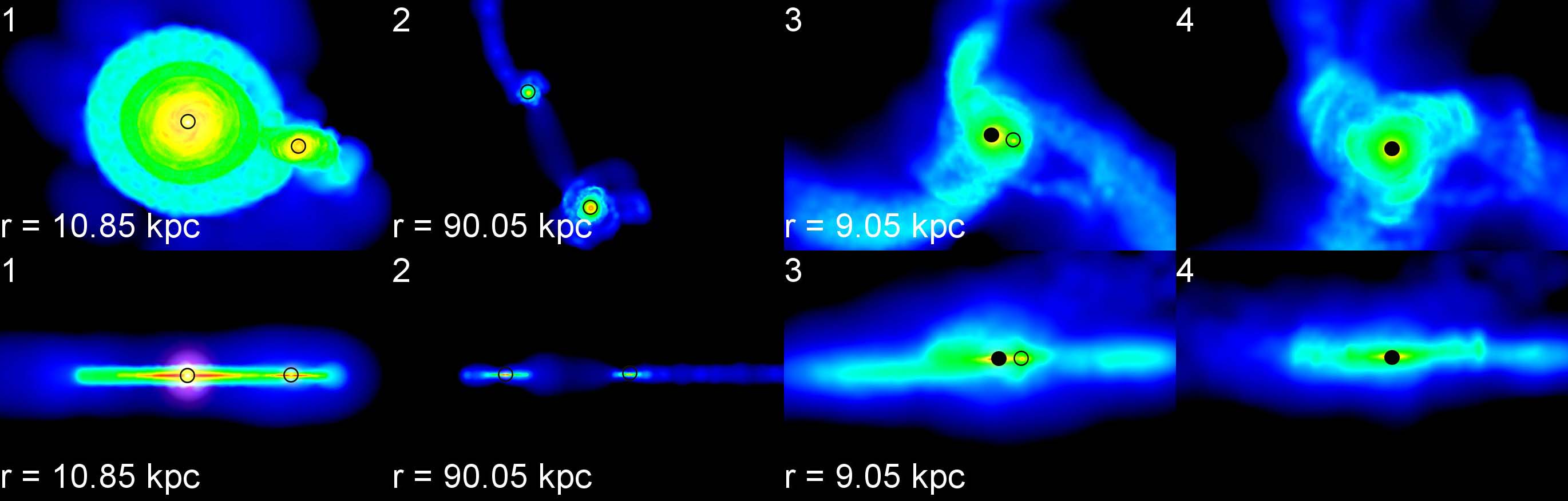}\\
    \includegraphics[width=0.9\textwidth]{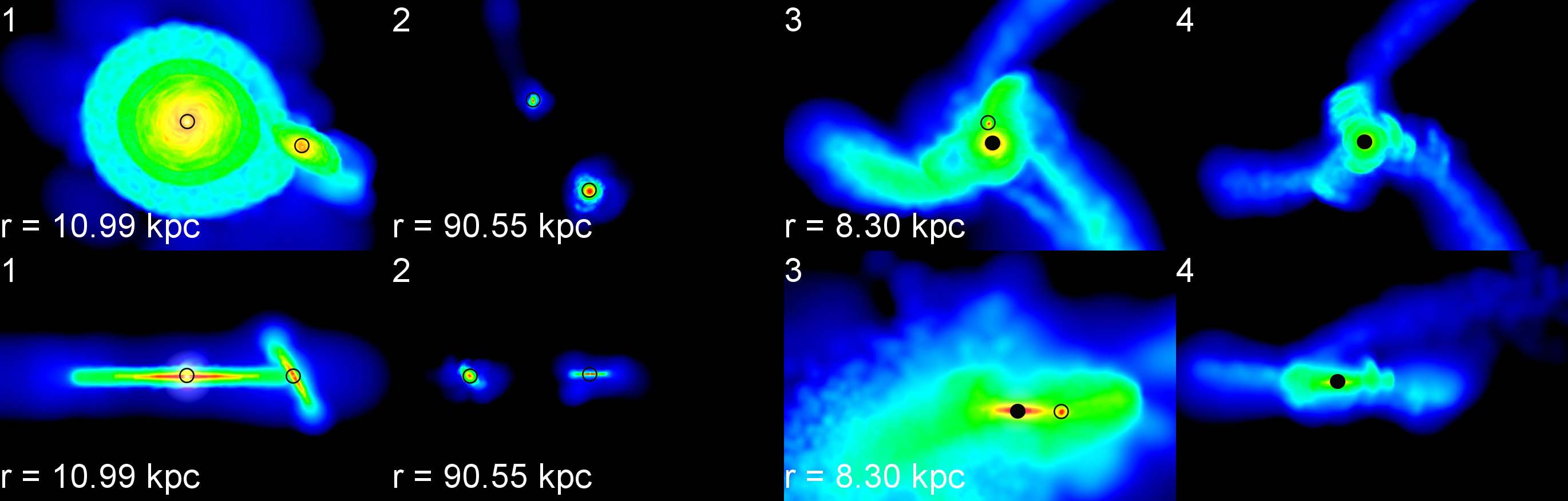}\\
        \includegraphics[width=0.9\textwidth]{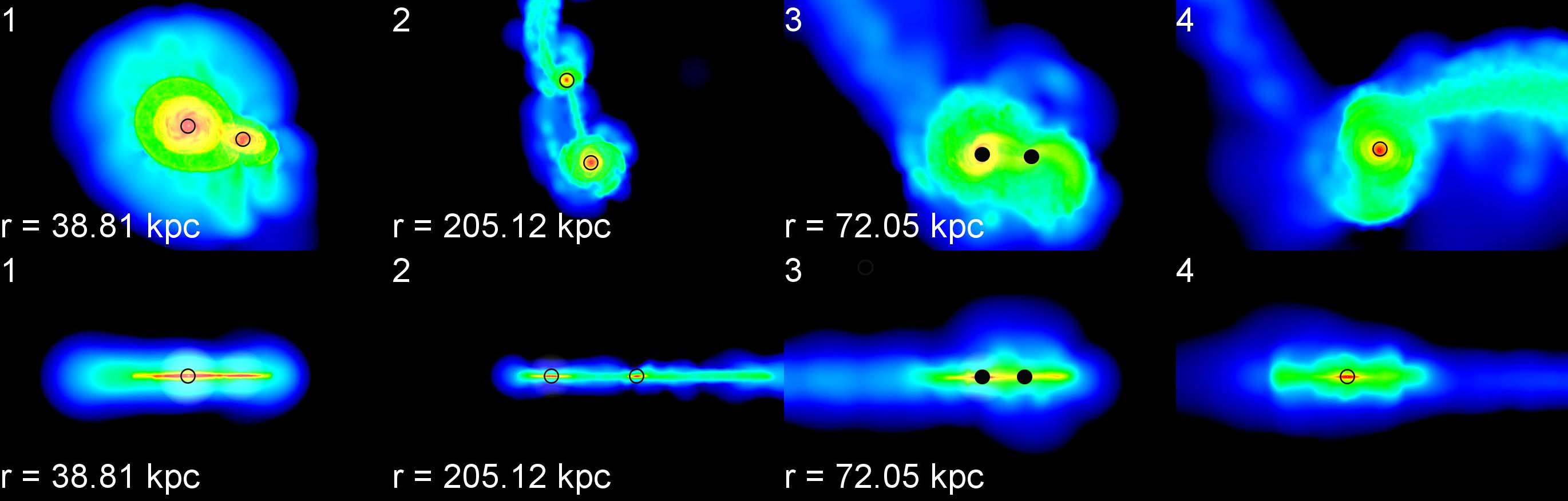}\\
    \includegraphics[width=0.9\textwidth]{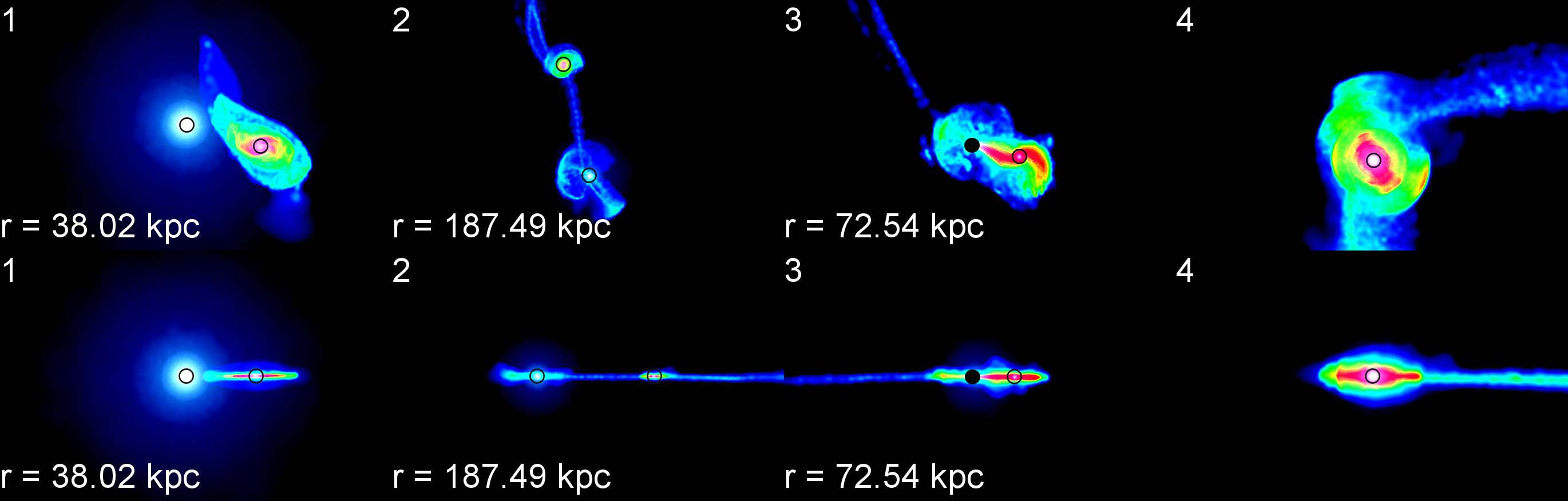}\\
    \includegraphics[width=0.9\textwidth]{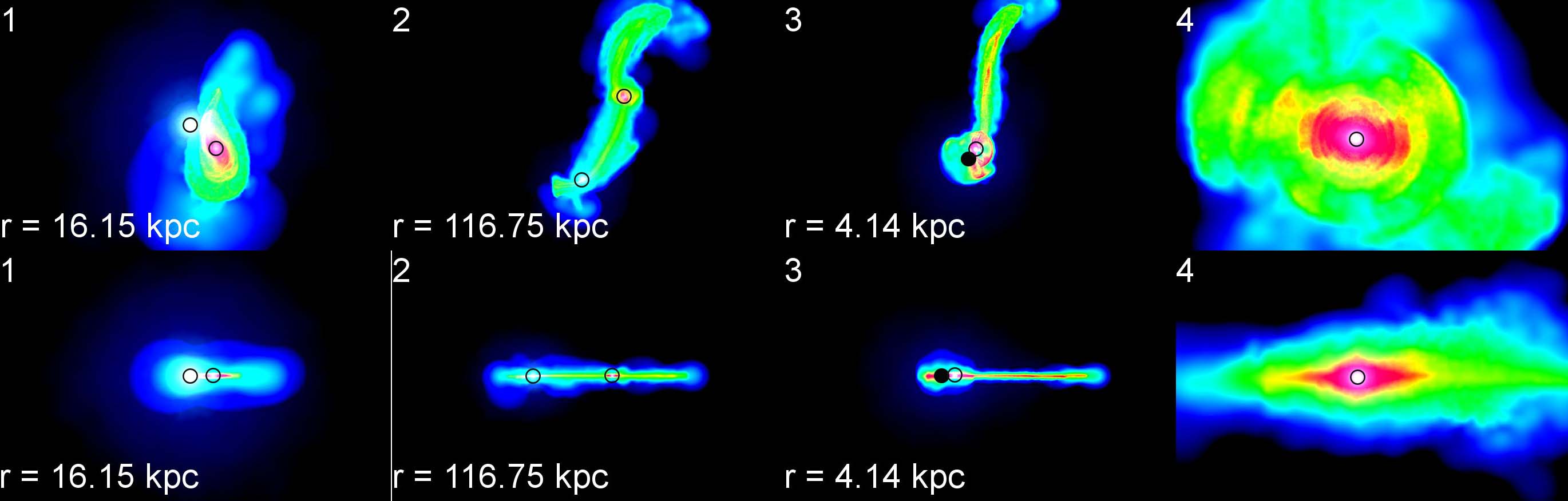}\\
    \caption{Snapshots of the simulations, legends are the same as  Figure~\ref{fig:galaxy1}, except that the simulations shown here are \textsf{q10f33ssz3}, \textsf{q10f33ssz3i}, \textsf{q5f33ssz1}, \textsf{q5f03esz1}, and \textsf{q5f03esz1p10}, respectively. }
    \label{fig:galaxy2}
\end{figure*}

Figures~\ref{fig:galaxy1} and \ref{fig:galaxy2} show the morphology of the merging galaxies in four different snapshots for each simulation: (1) the first pericentric passage, (2) the first apocentric passage after the first pericentric passage, (3) the time when one of the two nuclei is active, and (4) the last output of the simulation. In the two figures, each two rows show the four snapshots viewed by face-on (first row) and edge-on (second row) angles. After the first pericentric passage a tidal bridge appears, which is the channel for the material transportation between the two galaxies. In the cases that the secondary galaxy is colliding with inclination angle $i=45^{\circ}$ (\textsf{q5f33ssz3i} and \textsf{q10f33ssz3i}), a tidal tail outside the galactic plane can be clearly seen. The tidal tails and bridges are believed as the evidences of galaxy merger \citep{1995ApJ...438L..75M, 1996ApJ...471..115B, 2003AJ....126.1227K, 2007A&A...468...61D, 2007AJ....133..791S}

\begin{figure*}
    \centering
    \includegraphics[width=0.99\textwidth]{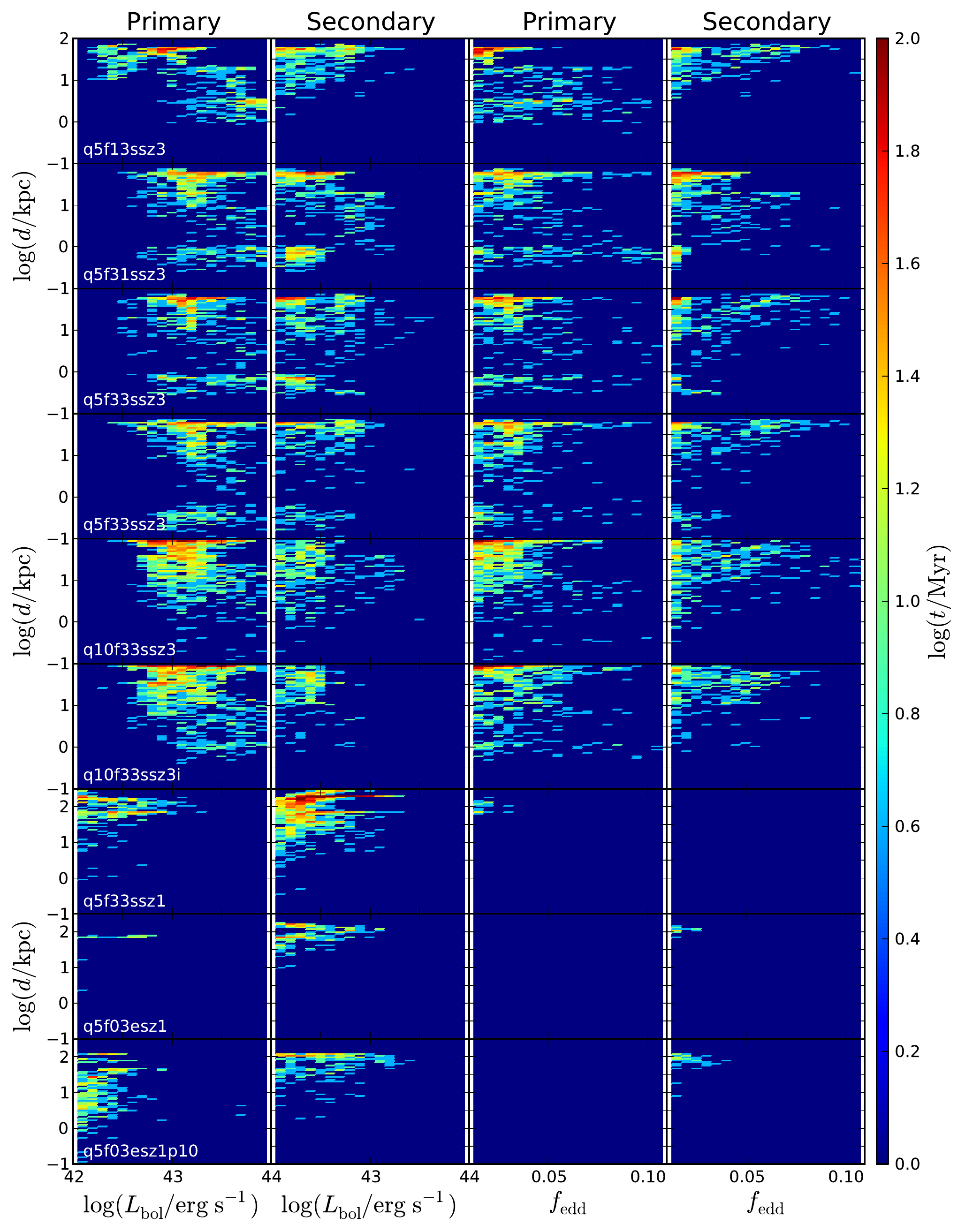}
    \caption{Duration time of the MBH activities at different separations ($\log(d/{\rm kpc})$) between the primary and secondary MBHs, bolometric luminosities (left two columns), and Eddington ratios (right two columns) for the primary (the first and third columns) and secondary (the second and fourth columns) galaxies. Simulations from top to bottom rows correspond to that shown in Table \ref{tab:ini_setup} from top to bottom rows, respectively. The right colorbar shows the exact duration time in different colors.}
    \label{fig:time_all}
\end{figure*}

For galaxy merging processes in different galaxy types and gas fractions, the MBH activities in the primary and secondary galaxy are triggered at different times, which rely on the gas contained in each galaxy. Figure~\ref{fig:time_all} shows the duration time at different separations, bolometric luminosities, and Eddington ratios for paired galaxies. Here the duration time means the observed timescale of an active nucleus at a given separation, bolometric luminosity, and Eddington ratio for each simulation run. From this figure we can obtain the following interpretation: 1) for those spiral-spiral minor mergers, the black hole activities have a dichotomy distributions: one peaks at larger separations ($\gtrsim 50$ kpc, and another one peaks at sub-kpc scale. 2) for the primary galaxies in the spiral-spiral minor mergers, following the mass increase of its central MBH, the $L_{\rm bol}$ increase at the sub-kpc activity peak, but the corresponding $f_{\rm Edd}$ are similar. 3) for the secondary galaxies in the spiral-spiral minor mergers, their $L_{\rm bol}$ at different separations have small oscillation, but their $f_{\rm Edd}$ decrease with smaller separations. 4) for the two elliptical-spiral mergers, if the two galaxies collide in closer encounter at the first pericentric passage (\textsf{q5f03esz1p10}, bottom row), the primary elliptical galaxy can capture the gas from the secondary spiral galaxy easier and its central MBH can accrete more gas and becomes more active than that of the larger encounter case (\textsf{q5f03esz1}, 8th row).

With the detected duration time of the two merging galaxies, we find that dAGN may emerge in several merging cases if using the lowest thresholds of $L_{\rm bol}=10^{43}{\rm erg~s^{-1}}$ or $f_{\rm Edd}$ as set in Section \ref{sec:method}. For the duration time of each MBH, Table \ref{tab:time_fraction} lists the results based on the two $L_{\rm bol}$ and two $f_{\rm Edd}$ thresholds, respectively. In all the minor merger cases, the secondary MBHs never have their $L_{\rm bol} \ge 10^{44}~{\rm erg~s^{-1}}$. For each simulation, we count the time duration when both MBHs are active and above the given $L_{\rm bol}$ or $f_{\rm Edd}$ thresholds, and list them in the row named `dAGN'. We find that not all the minor mergers can trigger observable dAGNs with significant time duration. Comparing \textsf{q5f13ssz3} with the other $z = 3$ cases, a system with the primary galaxy has lower gas fraction than the secondary galaxy can significantly decrease the detection rate of dAGNs. The last row `offset' of each simulation listed in Table \ref{tab:time_fraction} shows the time duration when the two nuclei reach the $L_{\rm bol}$ or $f_{\rm Edd}$ thresholds, and the secondary nucleus has larger luminosity or higher Eddington ratio than the primary galaxy, i.e. an oAGN system. From the `Offset' fraction detected under the $f_{\rm Edd}=0.01$ threshold, current nine merging systems only provide weak clues that oAGNs appear more frequently for those gas-rich mergers with the two galaxies have different gas fractions (e.g., \textsf{q5f13ssz3} and \textsf{q5f31ssz3}) than those gas-rich mergers with similar gas fractions (e.g., other spiral-spiral mergers) or elliptical-spiral mergers.

\begin{figure*}
    \centering
    \includegraphics[width=0.99\textwidth]{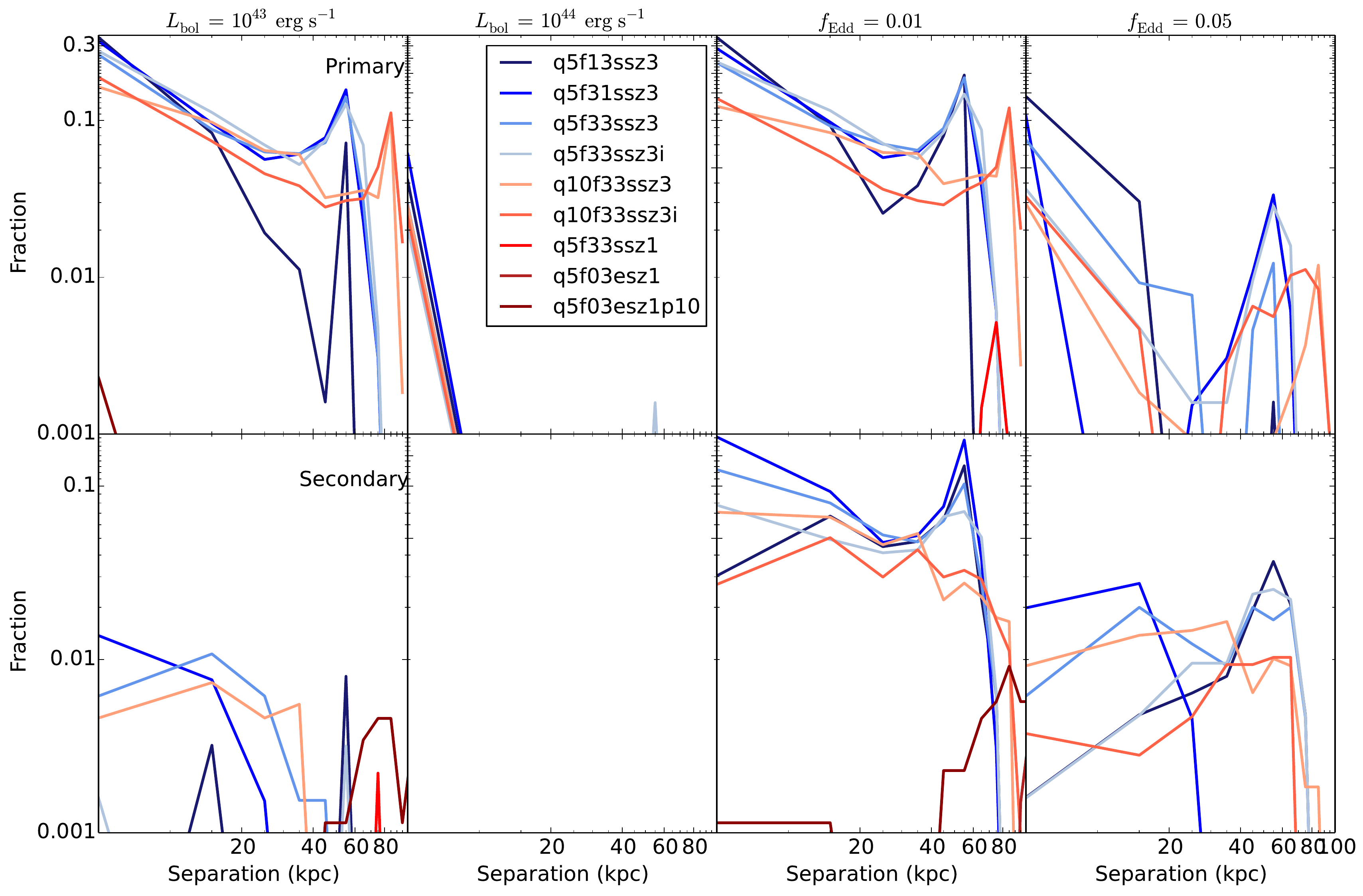}
    \caption{The AGN fractions at different separations detected in the two luminosity ($L_{\mathrm{bol}} = 10^{43}\ \mathrm{erg/s}$ for left column, $L_{\mathrm{bol}} = 10^{44}\ \mathrm{erg/s}$ for middle-left column) and two Eddington ratio ($f_{\mathrm{Edd}} = 0.01$ for middle-right column, $f_{\mathrm{Edd}} = 0.05$ for right column) thresholds for the primary (top row) and secondary (bottom row) MBHs. Line colors shown from blue to red show results of the 9 simulations listed from top to bottom in Table \ref{tab:ini_setup}.}
    \label{fig:time_frac}
\end{figure*}

Figure~\ref{fig:time_frac} summarizes the AGN fraction of the primary (top row) and secondary (bottom row) detected at different luminosity and Eddington ratio thresholds. The dichotomy distributions shown in Figure~\ref{fig:time_all} present clearer in Figure~\ref{fig:time_frac}. 
The peaks located at larger separation are caused by the strong interaction after the first pericentric passage (see details in Figure~\ref{fig:bhacc_z3_5}, Figure~\ref{fig:bhacc_z3_10} and Figure~\ref{fig:bhacc_z1}). It is not surprising that the time fraction reaches its maximum at small separation since the galaxy interaction induced nuclear activity. The AGN fraction in the three $z = 1$ simulations are hard to recognize because the two nuclei never reach to $\sim$ 100 pc in our simulation. None of the secondary nucleus can be more luminous than $L_{\mathrm{bol}} = 10^{44}\ \mathrm{erg\ s^{-1}}$.

\begin{table*}
    \centering
    \caption{AGN fractions in different thresholds of $L_{\rm bol}$ or $f_{\rm Edd}$.}
    \begin{tabular}{cc|cccc|cccc}
    \hline
    \multicolumn{2}{c|}{Run} & \multicolumn{2}{c}{$L_{\mathrm{bol}} = 10^{43}\ \mathrm{erg\ s^{-1}}$} & \multicolumn{2}{c|}{$L_{\mathrm{bol}} = 10^{44}\ \mathrm{erg\ s^{-1}}$} & \multicolumn{2}{c}{$f_{\mathrm{Edd}} = 0.01$} & \multicolumn{2}{c}{$f_{\mathrm{Edd}} = 0.05$}\\
    \cline{3-10}
    \multicolumn{2}{c|}{} & $t_{\mathrm{AGN}}$ & $t_{\mathrm{AGN}}/t_{\mathrm{tot}}$ & $t_{\mathrm{AGN}}$ & $t_{\mathrm{AGN}}/t_{\mathrm{tot}}$ & $t_{\mathrm{AGN}}$ & $t_{\mathrm{AGN}}/t_{\mathrm{tot}}$ & $t_{\mathrm{AGN}}$ & $t_{\mathrm{AGN}}/t_{\mathrm{tot}}$\\
    \hline
    \multirow{2}{*}{\textsf{q5f13ssz3}} & BH$_{1}$ & 1.31 & 0.52 & 0.06 & 0.03 & 1.92 & 0.77 & 0.42 & 0.17 \\
    & BH$_{2}$ & 0.01 & 0.004 & 0 & 0 & 1.03 & 0.41 & 0.22 & 0.09 \\
    & dAGN & 0 & 0 & 0 & 0 & 0.36 & 0.14 & 0 & 0 \\
    & Offset & 0.002 & 0.0008 & 0 & 0 & 0.37 & 0.15 & 0.03 & 0.01 \\
    \hline
    \multirow{2}{*}{\textsf{q5f31ssz3}} & BH$_{1}$ & 1.87 & 0.71 & 0.17 & 0.07 & 2.15 & 0.82 & 0.49 & 0.19 \\
    & BH$_{2}$ & 0.03 & 0.01 & 0 & 0 & 1.79 & 0.68 & 0.11 & 0.04\\
    & dAGN & 0 & 0 & 0 & 0 & 1.20 & 0.45 & 0.01 & 0.004 \\
    & Offset & 0 & 0 & 0 & 0 & 0.43 & 0.17 & 0.09 & 0.03\\
    \hline
    \multirow{2}{*}{\textsf{q5f33ssz3}} & BH$_{1}$ & 1.87 & 0.73 & 0.05 & 0.02 & 2.05 & 0.79 & 0.23 & 0.09\\
    & BH$_{2}$ & 0.03 & 0.01 & 0 & 0 & 1.31 & 0.51 & 0.27 & 0.11\\
    & dAGN & 0.01 & 0.004 & 0 & 0 & 1.13 & 0.44 & 0.007 & 0.003\\
    & Offset & 0 & 0 & 0 & 0 & 0.15 & 0.06 & 0.05 & 0.02\\
    \hline
    \multirow{2}{*}{\textsf{q5f33ssz3i}} & BH$_{1}$ & 2.00 & 0.80 & 0.05 & 0.02 & 2.03 & 0.80 & 0.12 & 0.05 \\
    & BH$_{2}$ & 0.003 & 0.001 & 0 & 0 & 1.02 & 0.41 & 0.28 & 0.11 \\
    & dAGN & 0.002 & 0.001 & 0 & 0 & 0.92 & 0.37 & 0 & 0\\
    & Offset & 0 & 0 & 0 & 0 & 0.11 & 0.04 & 0.02 & 0.008 \\
    \hline
    \multirow{2}{*}{\textsf{q10f33ssz3}} & BH$_{1}$ & 2.69 & 0.62 & 0.19 & 0.44 & 2.71 & 0.62 & 0.18 & 0.04 \\
    & BH$_{2}$ & 0.05 & 0.001 & 0 & 0 & 1.49 & 0.34 & 0.33 & 0.08 \\
    & dAGN & 0.02 & 0.005 & 0 & 0 & 1.33 & 0.31 & 0.004 & 0.001 \\
    & Offset & 0 & 0 & 0 & 0 & 0.28 & 0.06 & 0.22 & 0.05 \\
    \hline
    \multirow{2}{*}{\textsf{q10f33ssz3i}} & BH$_{1}$ & 2.45 & 0.57 & 0.11 & 0.03 & 2.39 & 0.56 & 0.28 & 0.07 \\
    & BH$_{2}$ & 0 & 0 & 0 & 0 & 1.16 & 0.27 & 0.26 & 0.06 \\
    & dAGN & 0 & 0 & 0 & 0 & 0.98 & 0.23 & 0 & 0\\
    & Offset & 0 & 0 & 0 & 0 & 0.15 & 0.04 & 0.005 & 0.001 \\
    \hline
    \multirow{2}{*}{\textsf{q5f33ssz1}} & BH$_{1}$ & 0.03 & 0.006 & 0 & 0 & 0 & 0 & 0 & 0 \\
    & BH$_{2}$ & 0.07 & 0.01 & 0 & 0 & 0.088 & 0.02 & 0 & 0 \\
    & dAGN & 0 & 0 & 0 & 0 & 0 & 0 & 0 & 0 \\
    & Offset & 0.05 & 0.01 & 0 & 0 & 0 & 0 & 0 & 0 \\
    \hline
    \multirow{2}{*}{\textsf{q5f03esz1}} & BH$_{1}$ & 0 & 0 & 0 & 0 & 0 & 0 & 0 & 0 \\
    & BH$_{2}$ & 0.005 & 0.001 & 0 & 0 & 0.07 & 0.01 & 0 & 0 \\
    & dAGN & 0 & 0 & 0 & 0 & 0 & 0 & 0 & 0  \\
    & Offset & 0.006 & 0.001 & 0 & 0 & 0.06 & 0.01 & 0 & 0 \\
    \hline
    \multirow{2}{*}{\textsf{q5f03esz1p10}} & BH$_{1}$ & 0.01 & 0.003 & 0 & 0 & 0 & 0 & 0 & 0 \\
    & BH$_{2}$ & 0.10 & 0.03 & 0 & 0 & 0.15 & 0.04 & 0 & 0 \\
    & dAGN & 0 & 0 & 0 & 0 & 0 & 0 & 0 & 0 \\
    & Offset & 0.05 & 0.01 & 0 & 0 & 0.14 & 0.04 & 0 & 0 \\
    \hline
    \end{tabular}
    \label{tab:time_fraction}
\end{table*}

\section{Conclusions}
\label{sec:conclusion}

We perform nine hydro-dynamical simulations with different settings of progenitor galaxies (mass ratio, gas fraction, starting redshift, and projected separation) to investigate how the nuclear activity can be triggered in galaxy minor mergers. We find that similar to galaxy major mergers, galaxy minor merger can trigger dAGNs but with a substantially smaller time duration (typically $\le 0.01$\,Gyr), more than an order of magnitude smaller than those by major merger (typically $\le 0.24$\,Gyr) \citep[e.g.,][]{2019SCPMA}. Minor Merger can also result in oAGNs with a time duration of $\le 0.22$\,Gyr. 

The Eddington ratios of the nuclear activities induced by minor mergers can hardly exceed $0.1$. As a comparison, those nuclear activities induced by major mergers can last for more than hundred million years with Eddington ratio larger than $0.1$. 

For all the simulations, the Eddington ratio of the primary galaxy increases after the first pericentric passage whatever the primary galaxy is initially gas poor ($f_{\rm gas} = 0.1$ in \textsf{q5f13ssz3} and $f_{\rm gas} = 0$ in \textsf{q5f03esz1} and \textsf{q5f03esz1p10}) or gas rich (the other six simulations), since the primary galaxy can always rob the gas from its companion. After the fourth pericentric passage, the Eddington ratio of the secondary galaxy decreases gradually after the gas is either consumed by star formation or captured by the primary galaxy during the interaction. In the dry-wet mergers at $z = 1$ (\textsf{q5f03esz1} and \textsf{q5f03esz1p10}), as the two galaxies approach to each other, the primary galaxy gradually robs more and more gas from the secondary galaxy to feed the central engine. However, the amount of gas is still inadequate to trigger the nuclear activity to a higher Eddington ratio.

The co-planar gas rich mergers generally trigger a relatively longer-lived AGN (including dAGN and oAGN) than that merging with an inclination angle (e.g., $45^\circ$ in our simulation), because the galaxy interaction tends to be strongest in the co-planar case, in which more gas can be transferred to the vicinity of the central MBH. From all the 9 runs we find that minor galaxy merger can trigger dAGN and oAGN systems but the time duration is relatively short compared to those gas rich major mergers \citep{2012ApJ...748L...7V, 2017MNRAS.469.4437C, 2019SCPMA}. Galaxy minor merger can be responsible for triggering nuclear activity as luminous as $L_{\mathrm{bol}} = 10^{44}\,\mathrm{erg/s}$ if both of the two progenitors are not too dry ($f_{\mathrm{gas}}$ should not be too smaller than $0.1$), especially at small BH separations.

\normalem
\begin{acknowledgements}
This work is supported by the National Key Program for Science and Technology Research and Development (Grant No. 2016YFA0400704), the National Natural Science Foundation of China (NSFC) under grant number 11690024 and 11873056, and the Strategic Priority Program of the Chinese Academy of Sciences (Grant No. XDB 23040100).
\end{acknowledgements}
  
\bibliographystyle{raa}

\begin{thebibliography}{}
\bibitem[Barnes \& Hernquist(1996)]{1996ApJ...471..115B} Barnes, J.~E., \& Hernquist, L.\ 1996, \apj, 471, 115 

\bibitem[Barrows et al.(2016)]{2016ApJ...829...37B} Barrows, R.~S., Comerford, J.~M., Greene, J.~E., \& Pooley, D.\ 2016, \apj, 829, 37 

\bibitem[Barrows et al.(2017)]{2017ApJ...838..129B} Barrows, R.~S., Comerford, J.~M., Greene, J.~E., \& Pooley, D.\ 2017, \apj, 838, 129 

\bibitem[Barrows et al.(2018)]{2018ApJ...869..154B} Barrows, R.~S., Comerford, J.~M., \& Greene, J.~E.\ 2018, \apj, 869, 154 

\bibitem[Begelman et al.(1980)]{1980Natur.287..307B} Begelman, M.~C., Blandford, R.~D., \& Rees, M.~J.\ 1980, \nat, 287, 307

\bibitem[Blecha et al.(2011)]{2011MNRAS.412.2154B} Blecha, L., Cox, T.~J., Loeb, A., \& Hernquist, L.\ 2011, \mnras, 412, 2154 

\bibitem[Blecha et al.(2013)]{2013MNRAS.429.2594B} Blecha, L., Loeb, A., \& Narayan, R.\ 2013, \mnras, 429, 2594 

\bibitem[Blecha et al.(2018)]{2018MNRAS.478.3056B} Blecha, L., Snyder, G.~F., Satyapal, S., \& Ellison, S.~L.\ 2018, \mnras, 478, 3056 

\bibitem[Bondi \& Hoyle(1944)]{1944MNRAS.104..273B} Bondi, H., \& Hoyle, F.\ 1944, \mnras, 104, 273 

\bibitem[Bondi(1952)]{1952MNRAS.112..195B} Bondi, H.\ 1952, \mnras, 112, 195 

\bibitem[Booth \& Schaye(2009)]{2009MNRAS.398...53B} Booth, C.~M., \& Schaye, J.\ 2009, \mnras, 398, 53

\bibitem[Campanelli et al.(2007)]{2007PhRvL..98w1102C} Campanelli, M., Lousto, C.~O., Zlochower, Y., \& Merritt, D.\ 2007, Physical Review Letters, 98, 231102 

\bibitem[Capelo et al.(2015)]{2015MNRAS.447.2123C} Capelo, P.~R., Volonteri, M., Dotti, M., et al.\ 2015, \mnras, 447, 2123 

\bibitem[Capelo et al.(2017)]{2017MNRAS.469.4437C} Capelo, P.~R., Dotti, M., Volonteri, M., et al.\ 2017, \mnras, 469, 4437 

\bibitem[Cisternas et al.(2011)]{2011ApJ...726...57C} Cisternas, M., Jahnke, K., Inskip, K.~J., et al.\ 2011, \apj, 726, 57

\bibitem[Comerford et al.(2009)]{2009ApJ...698..956C} Comerford, J.~M., Gerke, B.~F., Newman, J.~A., et al.\ 2009, \apj, 698, 956

\bibitem[Comerford et al.(2011)]{2011ApJ...737L..19C} Comerford, J.~M., Pooley, D., Gerke, B.~F., \& Madejski, G.~M.\ 2011, \apjl, 737, L19

\bibitem[Comerford et al.(2012)]{2012ApJ...753...42C} Comerford, J.~M., Gerke, B.~F., Stern, D., et al.\ 2012, \apj, 753, 42 

\bibitem[Comerford \& Greene(2014)]{2014ApJ...789..112C} Comerford, J.~M., \& Greene, J.~E.\ 2014, \apj, 789, 112 

\bibitem[Comerford et al.(2015)]{2015ApJ...806..219C} Comerford, J.~M., Pooley, D., Barrows, R.~S., et al.\ 2015, \apj, 806, 219 

\bibitem[Comerford et al.(2017)]{2017ApJ...847...41C} Comerford, J.~M., Barrows, R.~S., Greene, J.~E., \& Pooley, D.\ 2017, \apj, 847, 41 

\bibitem[Comerford et al.(2018)]{2018ApJ...867...66C} Comerford, J.~M., Nevin, R., Stemo, A., et al.\ 2018, \apj, 867, 66

\bibitem[Conselice(2014)]{2014ARA&A..52..291C} Conselice, C.~J.\ 2014, \araa, 52, 291 

\bibitem[Di Matteo et al.(2005)]{2005Natur.433..604D} Di Matteo, T., Springel, V., \& Hernquist, L.\ 2005, \nat, 433, 604 

\bibitem[Di Matteo et al.(2007)]{2007A&A...468...61D} Di Matteo, P., Combes, F., Melchior, A.-L., \& Semelin, B.\ 2007, \aap, 468, 61 

\bibitem[Donley et al.(2018)]{2018ApJ...853...63D} Donley, J.~L., Kartaltepe, J., Kocevski, D., et al.\ 2018, \apj, 853, 63

\bibitem[Ellison et al.(2011)]{2011MNRAS.418.2043E} Ellison, S.~L., Patton, D.~R., Mendel, J.~T., \& Scudder, J.~M.\ 2011, \mnras, 418, 2043

\bibitem[Fu et al.(2011a)]{2011ApJ...733..103F} Fu, H., Myers, A.~D., Djorgovski, S.~G., \& Yan, L.\ 2011, \apj, 733, 103 

\bibitem[Fu et al.(2011b)]{2011ApJ...740L..44F} Fu, H., Zhang, Z.-Y., Assef, R.~J., et al.\ 2011, \apjl, 740, L44 

\bibitem[Fu et al.(2012)]{2012ApJ...745...67F} Fu, H., Yan, L., Myers, A.~D., et al.\ 2012, \apj, 745, 67

\bibitem[Fu et al.(2018)]{2018ApJ...856...93F} Fu, H., Steffen, J.~L., Gross, A.~C., et al.\ 2018, \apj, 856, 93 

\bibitem[Frey et al.(2012)]{2012MNRAS.425.1185F} Frey, S., Paragi, Z., An, T., \& Gab{\'a}nyi, K.~{\'E}.\ 2012, \mnras, 425, 1185

\bibitem[Ge et al.(2012)]{2012ApJS..201...31G} Ge, J.-Q., Hu, C., Wang, J.-M., Bai, J.-M., \& Zhang, S.\ 2012, \apjs, 201, 31

\bibitem[Goulding et al.(2018)]{2018PASJ...70S..37G} Goulding, A.~D., Greene, J.~E., Bezanson, R., et al.\ 2018, \pasj, 70, S37

\bibitem[Hayward et al.(2014)]{2014MNRAS.442.1992H} Hayward, C.~C., Torrey, P., Springel, V., Hernquist, L., \& Vogelsberger, M.\ 2014, \mnras, 442, 1992 

\bibitem[Hernquist(1989)]{1989Natur.340..687H} Hernquist, L.\ 1989, \nat, 340, 687 

\bibitem[Hernquist(1990)]{1990ApJ...356..359H} Hernquist, L.\ 1990, \apj, 356, 359 

\bibitem[Hewlett et al.(2017)]{2017MNRAS.470..755H} Hewlett, T., Villforth, C., Wild, V., et al.\ 2017, \mnras, 470, 755

\bibitem[Hong et al.(2015)]{2015ApJ...804...34H} Hong, J., Im, M., Kim, M., \& Ho, L.~C.\ 2015, \apj, 804, 34

\bibitem[Hopkins et al.(2008)]{2008ApJS..175..356H} Hopkins, P.~F., Hernquist, L., Cox, T.~J., \& Kere{\v s}, D.\ 2008, \apjs, 175, 356 

\bibitem[Hoyle \& Lyttleton(1939)]{1939PCPS...35..405H} Hoyle, F., \& Lyttleton, R.~A.\ 1939, Proceedings of the Cambridge Philosophical Society, 35, 405 

\bibitem[Ilbert et al.(2010)]{2010ApJ...709..644I} Ilbert, O., Salvato, M., Le Floc'h, E., et al.\ 2010, \apj, 709, 644 

\bibitem[Johansson et al.(2009)]{2009ApJ...690..802J} Johansson, P.~H., Naab, T., \& Burkert, A.\ 2009, \apj, 690, 802

\bibitem[Knierman et al.(2003)]{2003AJ....126.1227K} Knierman, K.~A., Gallagher, S.~C., Charlton, J.~C., et al.\ 2003, \aj, 126, 1227 

\bibitem[Kocevski et al.(2012)]{2012ApJ...744..148K} Kocevski, D.~D., Faber, S.~M., Mozena, M., et al.\ 2012, \apj, 744, 148 

\bibitem[Komossa et al.(2003)]{2003ApJ...582L..15K} Komossa, S., Burwitz, V., Hasinger, G., et al.\ 2003, \apjl, 582, L15 

\bibitem[Kormendy \& Richstone(1995)]{1995ARA&A..33..581K} Kormendy, J., \& Richstone, D.\ 1995, \araa, 33, 581 

\bibitem[Kormendy \& Ho(2013)]{2013ARA&A..51..511K} Kormendy, J., \& Ho, L.~C.\ 2013, \araa, 51, 511

\bibitem[Koss et al.(2011)]{2011ApJ...735L..42K} Koss, M., Mushotzky, R., Treister, E., et al.\ 2011, \apjl, 735, L42

\bibitem[Koss et al.(2012)]{2012ApJ...746L..22K} Koss, M., Mushotzky, R., Treister, E., et al.\ 2012, \apjl, 746, L22

\bibitem[Koss et al.(2018)]{2018Natur.563..214K} Koss, M.~J., Blecha, L., Bernhard, P., et al.\ 2018, \nat, 563, 214 

\bibitem[Liu et al.(2010a)]{2010ApJ...708..427L} Liu, X., Shen, Y., Strauss, M.~A., \& Greene, J.~E.\ 2010, \apj, 708, 427

\bibitem[Liu et al.(2010b)]{2010ApJ...715L..30L} Liu, X., Greene, J.~E., Shen, Y., \& Strauss, M.~A.\ 2010, \apjl, 715, L30

\bibitem[Lofthouse et al.(2017)]{2017MNRAS.465.2895L} Lofthouse, E.~K., Kaviraj, S., Conselice, C.~J., Mortlock, A., \& Hartley, W.\ 2017, \mnras, 465, 2895


\bibitem[Madau \& Quataert(2004)]{2004ApJ...606L..17M} Madau, P., \& Quataert, E.\ 2004, \apjl, 606, L17 

\bibitem[Magorrian et al.(1998)]{1998AJ....115.2285M} Magorrian, J., Tremaine, S., Richstone, D., et al.\ 1998, \aj, 115, 2285 

\bibitem[Marconi \& Hunt(2003)]{2003ApJ...589L..21M} Marconi, A., \& Hunt, L.~K.\ 2003, \apjl, 589, L21 

\bibitem[Menci et al.(2014)]{2014A&A...569A..37M} Menci, N., Gatti, M., Fiore, F., \& Lamastra, A.\ 2014, \aap, 569, A37

\bibitem[Mihos(1995)]{1995ApJ...438L..75M} Mihos, J.~C.\ 1995, \apjl, 438, L75 

\bibitem[M{\"u}ller-S{\'a}nchez et al.(2015)]{2015ApJ...813..103M} M{\"u}ller-S{\'a}nchez, F., Comerford, J.~M., Nevin, R., et al.\ 2015, \apj, 813, 103 

\bibitem[M{\"u}ller-S{\'a}nchez et al.(2016)]{2016ApJ...830...50M} M{\"u}ller-S{\'a}nchez, F., Comerford, J., Stern, D., \& Harrison, F.~A.\ 2016, \apj, 830, 50 

\bibitem[Nagamine et al.(2004)]{2004MNRAS.348..435N} Nagamine, K., Springel, V., \& Hernquist, L.\ 2004, \mnras, 348, 435

\bibitem[Negri \& Volonteri(2017)]{2017MNRAS.467.3475N} Negri, A., \& Volonteri, M.\ 2017, \mnras, 467, 3475

\bibitem[Rosas-Guevara et al.(2015)]{2015MNRAS.454.1038R} Rosas-Guevara, Y.~M., Bower, R.~G., Schaye, J., et al.\ 2015, \mnras, 454, 1038


\bibitem[Satyapal et al.(2014)]{2014MNRAS.441.1297S} Satyapal, S., Ellison, S.~L., McAlpine, W., et al.\ 2014, \mnras, 441, 1297

\bibitem[Secrest et al.(2017)]{2017ApJ...836..183S} Secrest, N.~J., Schmitt, H.~R., Blecha, L., Rothberg, B., \& Fischer, J.\ 2017, \apj, 836, 183 

\bibitem[Skipper \& Browne(2018)]{2018MNRAS.475.5179S} Skipper, C.~J., \& Browne, I.~W.~A.\ 2018, \mnras, 475, 5179 

\bibitem[Smith et al.(2007)]{2007AJ....133..791S} Smith, B.~J., Struck, C., Hancock, M., et al.\ 2007, \aj, 133, 791 

\bibitem[Sparre \& Springel(2017)]{2017MNRAS.470.3946S} Sparre, M., \& Springel, V.\ 2017, \mnras, 470, 3946

\bibitem[Springel \& Hernquist(2003)]{2003MNRAS.339..289S} Springel, V., \& Hernquist, L.\ 2003, \mnras, 339, 289 

\bibitem[Springel(2005)]{2005MNRAS.364.1105S} Springel, V.\ 2005, \mnras, 364, 1105 

\bibitem[Springel et al.(2005)]{2005MNRAS.361..776S} Springel, V., Di Matteo, T., \& Hernquist, L.\ 2005, \mnras, 361, 776

\bibitem[Springel \& Hernquist(2005)]{2005ApJ...622L...9S} Springel, V., \& Hernquist, L.\ 2005, \apjl, 622, L9

\bibitem[Steinborn et al.(2016)]{2016MNRAS.458.1013S} Steinborn, L.~K., Dolag, K., Comerford, J.~M., et al.\ 2016, \mnras, 458, 1013 

\bibitem[Thompson et al.(2014)]{2014ApJ...780..145T} Thompson, R., Nagamine, K., Jaacks, J., \& Choi, J.-H.\ 2014, \apj, 780, 145

\bibitem[Treister et al.(2012)]{2012ApJ...758L..39T} Treister, E., Schawinski, K., Urry, C.~M., \& Simmons, B.~D.\ 2012, \apjl, 758, L39

\bibitem[Tremaine et al.(2002)]{2002ApJ...574..740T} Tremaine, S., Gebhardt, K., Bender, R., et al.\ 2002, \apj, 574, 740

\bibitem[Van Wassenhove et al.(2012)]{2012ApJ...748L...7V} Van Wassenhove, S., Volonteri, M., Mayer, L., et al.\ 2012, \apjl, 748, L7 

\bibitem[Villforth et al.(2019)]{2019MNRAS.483.2441V} Villforth, C., Herbst, H., Hamann, F., et al.\ 2019, \mnras, 483, 2441

\bibitem[Wang et al.(2009)]{2009ApJ...705L..76W} Wang, J.-M., Chen, Y.-M., Hu, C., et al.\ 2009, \apjl, 705, L76

\bibitem[Wang et al.(2019)]{2019MNRAS.482.1889W} Wang, M.-X., Luo, A.-L., Song, Y.-H., et al.\ 2019, \mnras, 482, 1889 

\bibitem[Xu \& Komossa(2009)]{2009ApJ...705L..20X} Xu, D., \& Komossa, S.\ 2009, \apjl, 705, L20 

\bibitem[Yang(2019)]{2019SCPMA} Yang, C., Ge, J., \& Lu, Y. \ 2019, Science China Physics, Mechanics, and Astronomy, 62, 129511

\bibitem[Yu(2002)]{2002MNRAS.331..935Y} Yu, Q.\ 2002, \mnras, 331, 935 

\bibitem[Zhang \& Feng(2016)]{2016MNRAS.457.3878Z} Zhang, X.-G., \& Feng, L.-L.\ 2016, \mnras, 457, 3878

\bibitem[Zhou et al.(2004)]{2004ApJ...604L..33Z} Zhou, H., Wang, T., Zhang, X., Dong, X., \& Li, C.\ 2004, \apjl, 604, L33 

\end{thebibliography}

\end{document}